# Velocity-selective spectroscopy measurements of Rydberg fine structure states in a hot vapor cell

**Jun He,**[1,2,3] **Dongliang Pei,**[1] **Jieying Wang,**[1] **Junmin Wang**[1,2,3]*

[1]*Institute of Opto-Electronics, Shanxi University, Tai Yuan 030006*
[2]*State Key Laboratory of Quantum Optics and Quantum Optics Devices, Shanxi University, Tai Yuan 030006*
[3]*Collaborative Innovation Center of Extreme Optics, Tai Yuan 030006*
*\*Corresponding author:* Email: wwjjmm@sxu.edu.cn

**A velocity-selective spectroscopy technique for studying the spectra from hot Rydberg gases is presented. This method provides high-resolution measurement of the spectrum interval. Based on this method, the velocity-selective hyperfine splitting of intermediate states is measured, as well as the Doppler-free fine-structure splitting of the Rydberg states via two-photon Rydberg excitation in a room-temperature $^{133}$Cs vapor cell. The experiment data and theoretical predictions show excellent agreement.**

*OCIS codes: (020.1670) Coherent optical effects; (020.5780) Rydberg states; (300.6210) Spectroscopy, atomic; (020.4180) Multiphoton processes; (020.2930) Hyperfine structure.*

The Rydberg atom possesses one electron in a very high excited state, exemplifying a perfectly quantum system with a macroscopic size. Rydberg atoms have recently received considerable attention because of their large polarizability, strong dipole–dipole coupling and long lifetime. The large polarizability indicates a sensitivity to electric fields, which makes Rydberg atoms very promising systems for producing high-precision field sensors, similar to the way in which the realization of nonlinear optical effects was used to implement non-destructive measurements of single photons [1-4]. The strong dipole–dipole interaction renders Rydberg atoms promising candidates for the implementation of protocols realizing quantum gates or efficient multi-particle entanglement for quantum information processing [5-7]. For quantum information applications, a nondestructive detection of the Rydberg state is preferable. Electromagnetically-induced transparency (EIT) is a quantum destructive interference effect with a narrow, subnatural linewidth [8, 9], which provides a detection technique with no absorption and allows dissipation-free sensing of the desired atomic resonance. Therefore, EIT can be used to control the transmission properties of an atomic ensemble with very weak switching fields. The subnatural linewidth EIT effect thus makes it possible to precisely measure the Rydberg series of energy levels. In recent experiments, EIT has been used to observe the Rydberg dipole blockade in cold ensembles of atoms, and it has been proposed to directly observe the dipole blockade using EIT as well as vector microwave electrometry [10-12].

Great progress has also been made toward a ladder configuration of the Rydberg states in room-temperature vapor cells. Experimental efforts are being made to advance a precise optical spectrum technique to resolve the Rydberg states [9, 13-18]. Not only is this a very effective method for optical detection of the Rydberg states, but it is also a good way to measure optical non-linearity owing to the strong interaction between Rydberg atoms. The EIT signal intensity is limited primarily by the weak transition probability amplitude of Rydberg states, while the signal noise arises from the phase and intensity noise of the laser.

Using atom systems to make absolute energy shift measurements is difficult, and especially for large-scale changes in the principal quantum number. It should be emphasized that the energy shift of the strong interaction is not directly sensitive to the absolute frequency, and it is therefore not necessary to work with high accuracy to distinguish the absolute energy. In this work, we present experimental results for a ladder system EIT involving highly-excited Rydberg states. Based on a radio-frequency (RF) modulation technique, we measure the velocity dependence of the hyperfine splitting of intermediate states and the Doppler-free splitting of Rydberg states in a room-temperature $^{133}$Cs vapor cell. Using velocity-selective spectroscopy combined with a RF modulation technique to measure the relative energy shift requires only an interval of the spectrum that is accurately distinguishable, where the measurement precision can be made sensitive to a RF and limited to the EIT linewidth.

An external cavity diode laser with a wavelength of $\lambda_p$=852 nm was used as the probing laser, with a typical linewidth in the megahertz scale. Infrared light from a diode laser with a wavelength of 1018 nm was amplified up to 5 W by the fiber amplifier, and the output beam was frequency doubled in a periodically-poled KTP crystal (KTiOPO$_4$) to produce a $\lambda_c$=509 nm laser. The 852 and 509 nm-wavelength beams were then circularly polarized and overlapped in the cell. The cell was tens of millimeters in size to match the Raleigh length of the focused beams (~100 μm). The 852 nm-wavelength laser was locked to one of the $6S_{1/2} \rightarrow 6P_{3/2}$ hyperfine transitions via saturated absorption spectroscopy, and the laser wavelength was modulated by a waveguide phase-type electro-optic modulator. The 509 nm-wavelength enhanced doubling cavity was stabilized to the 1018 nm laser using the Pound-Drever-Hall RF modulation sideband method, and was fed back to a piezoelectric transducer (PZT). The entire system used ultra-stable mirror mounts to suppress noise. The model of the RF function generator (E8257D, Agilent Technologies) was locked to a rubidium clock (PRS10, Stanford Research Systems).

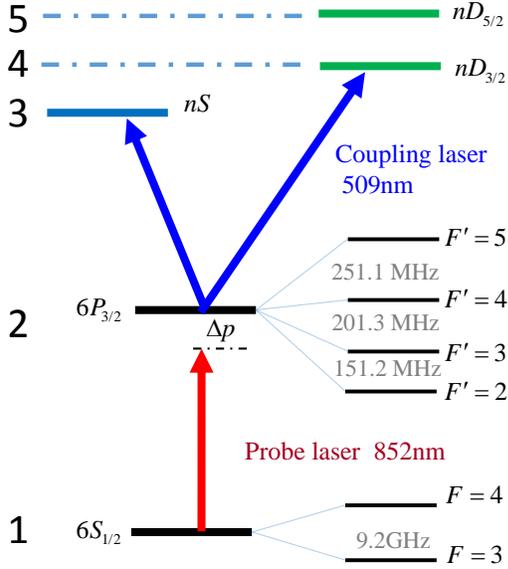

Fig. 1. Energy level schematic of the cascade-type electromagnetically-induced transparency (EIT) of $^{133}$Cs. The $\lambda_p$=852 nm probe laser is frequency-stabilized to the hyperfine transition to measure the absorption on the $6S_{1/2}(F=4) \rightarrow 6P_{3/2}(F'=5)$ transition, while the $\lambda_c$=509 nm laser beam couples to the $6P_{3/2}(F) \rightarrow nS_{1/2}$ transition.

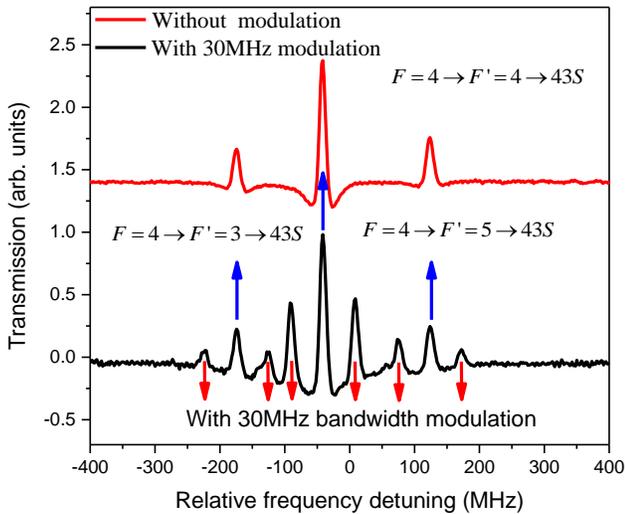

Fig. 2. Experimental EIT spectra across the principle quantum number $n$=43S transition with and without RF bandwidth modulation.

For a three-level ladder-type EIT, the matrix element $\rho_{21}$ can be derived by solving the steady-state optical Bloch equations using a perturbation approximation under the weak probe field regime. Taking into account the Doppler effect and the Boltzmann velocity distribution, the EIT spectral line shape when applying an approximate expression for the susceptibility can be written as
$\chi(\upsilon)d\upsilon =$

$$\frac{4i\hbar g_{21}^2/\varepsilon_0}{\gamma_{21}-i\Delta_p-i\frac{\omega_p}{c}\upsilon+\frac{\Omega_c^2/4}{\gamma_{31}-i(\Delta_p+\Delta_c)-i(\omega_p \pm \omega_c)\upsilon/c}}N(\upsilon)d\upsilon \quad (1)$$

where $c$ is the speed of light; $\upsilon$ is the atomic velocity along the direction of the probe beam ("+" ("−") for the co- (counter-) propagation configuration); $\Omega_p$ and $\Omega_c$ are the probe and coupling laser Rabi frequencies, respectively; $\Delta_p = \omega_p - \omega_{12}$ and $\Delta_c = \omega_c - \omega_{23}$ are the probe and coupling laser detunings, respectively; $N(\upsilon)d\upsilon = (N_0 u^{-1}\pi^{-1/2})e^{-x}d\upsilon$ (where $x = \upsilon^2/u^2$) is the number density of atoms with velocity $\upsilon$; $u = \sqrt{kT/m}$, where $k$ is the Boltzmann constant, $T$ is temperature, $\hbar$ is Planck's constant and $m$ is the mass of the $^{133}$Cs atom; $\gamma_{21}$ and $\gamma_{31}$ are the natural widths of the intermediate and upper states, respectively; and $g_{21}$ is the dipole moment matrix element.

Figure 1 shows the energy levels associated with the ladder-type Rydberg EIT. The EIT signal is observed by scanning the frequency of the $\lambda_c$=509 nm coupling laser while locking the frequency of the $\lambda_p$=852 nm probe laser. The background-free EIT spectra with its high signal-to-noise ratios is a benefit for the low-noise laser system and the very weak laser power, which decreases the intensity noise and phase noise. Figure 2 shows the EIT signal of the hyperfine splitting of the intermediate state originating from non-stationary atoms at room temperature. All EIT signals of the intermediate states are observed because of the existence of different velocity classes of atoms. Owing to the Doppler mismatch, the hyperfine splitting of the 5P$_{3/2}$ states scale as $\Delta_c = (1-\lambda_p/\lambda_c)\Delta_p$. For atoms with velocity $\upsilon$ moving in the same direction as the probe field, the detuning of the probe laser is $\Delta'_p \rightarrow -\omega_p \cdot \upsilon/c$ and that of the coupling laser is $\Delta'_c = \Delta'_p \cdot \omega_c/\omega_p$. Setting the transition frequency of $6S_{1/2}(F=4) \rightarrow 6P_{3/2}(F'=4) \rightarrow 42S$ as the reference frequency, the atoms with specific velocity groups $\upsilon_{4-4}$ and $\upsilon_{4-3}$ (where $\upsilon_{4-4}/c$=251.0 MHz and $\upsilon_{4-3}/c$=452.2 MHz) result in a resonance with other hyperfine states of $F=4 \rightarrow F'=4$ and $F=4 \rightarrow F'=3$, respectively. If the frequency of the 509 nm-wavelength coupling field matches these specific atoms groups to the levels of 43S, the Rydberg EIT signal can be detected, as shown in Figure 2. The relative intensity of the transparency peaks under multi-intermediate levels of the ladder-type EIT depends on the population in the intermediate states.

The RF modulation combined with a velocity selection scheme can further tune the EIT peaks via Doppler shifts, which depend on the Boltzmann factor $\exp(\hbar\Delta/k_BT)$ and can be used to measure the peak splitting. The spectral resolution is limited only by the EIT linewidth of the optical detection.

The signal intensity increases with the coupling and probe laser power, while the linewidth also increases. In the limit of a relatively high power for the coupling and probe field, the observed linewidth is ~9 MHz; and the broadened linewidth from 5.2 MHz for $\lambda_p$=852 nm to 9 MHz for $\lambda_c$=509 nm for the near-resonant atoms is owing to the wavelength mismatch broadening. The dips in the enhanced absorption that is observed on both wings of the EIT signal originate from the residual Doppler effect owing to the wavelength mismatch of the probe and the coupling lasers.

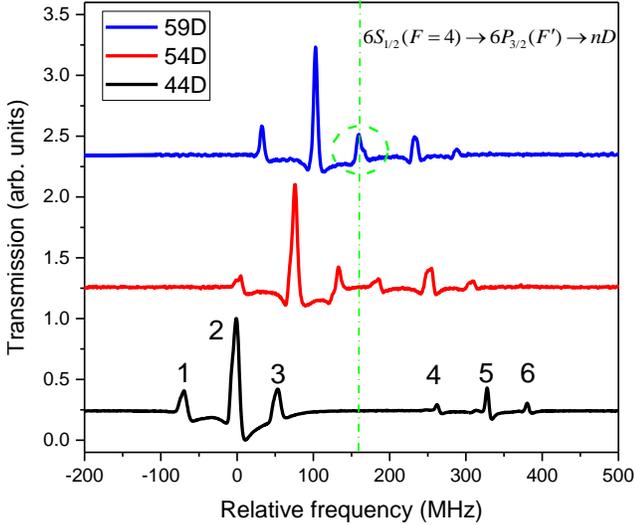

Fig. 3. Typical electromagnetically-induced transparency (EIT) spectra of the $6S_{1/2}(F=4) \to 6P_{3/2}(F') \to nD$ hyperfine splitting of intermediate states and fine splitting ($D_{3/2}$ and $D_{5/2}$) of Rydberg states. Because of the Rydberg splitting, the $D_{3/2}$ and $D_{5/2}$ states scale with $\Delta_c = (1 - \lambda_p / \lambda_c)\Delta_p$. The peaks labeled 1–3 respectively represent $F=4 \to F'=(3,4,5) \to nD_{5/2}$, and the peaks labeled 4–6 respectively represent $F=4 \to F'=(3,4,5) \to nD_{3/2}$. The horizontal coordinate is calibrated using the modulation sidebands of the $\lambda$=509 nm laser.

Figure 3 shows typical nD Rydberg states spectra, where the signal splitting decreases with the principal quantum number, $n$. For a cesium atom at room temperature, the peak of $6S_{1/2}(F=4) \to 6P_{3/2}(F'=5) \to 59D_{5/2}$ overlaps with the peak of $6S_{1/2}(F=4) \to 6P_{3/2}(F'=3) \to 59D_{3/2}$, which arises from the Doppler-free nature of the fine-structure splitting of the Rydberg state and the velocity-dependent nature of the hyperfine splitting of $6P_{3/2}(F')$. In the condition wherein $n$=59, the peaks of $6S_{1/2}(F=4) \to 6P_{3/2}(F'=5) \to 59D_{5/2}$ and $6S_{1/2}(F=4) \to 6P_{3/2}(F'=3) \to 59D_{3/2}$ overlap. The strong interaction of the Rydberg states causes their fine-structure peaks to shift. The RF-modulated EIT can distinguish the interval changes of the peaks, where the resolution is only limited by the linewidth of the optical detection. The RF modulation combined with a velocity selection scheme can further tune the EIT peaks via Doppler shifts, which can be used to measure the peak splitting.

Figure 4 shows the dependence of the fine structure and hyperfine structure on the principal quantum number, $n$. The horizontal coordinate is calibrated by the modulation sidebands of the $\lambda$=509 nm laser. In the case of the hyperfine structure of the intermediate states, the spectral splitting only depends upon a Doppler effect that is an order of magnitude larger than the principal quantum number $n$, and the spectrum for small $n$ is dominated by the $D_{3/2}$ and $D_{5/2}$ states. In the case of the fine structure of the Rydberg states, the spectral splitting depends sensitively on $n$ and appears to overlap with the spectra of other states as the principal quantum number increases. Herein, we use velocity-selective spectroscopy with RF modulation as a reference to calibrate the Rydberg fine-structure states in the hot vapor cell, where the RF frequency precision is smaller than a hertz for long time scales and the EIT linewidth is smaller than 9 MHz. The dominant deviations existing between the experimental and calculated results may arise from the nonlinear correspondence of the PZT while scanning the 509 nm-wavelength cavity.

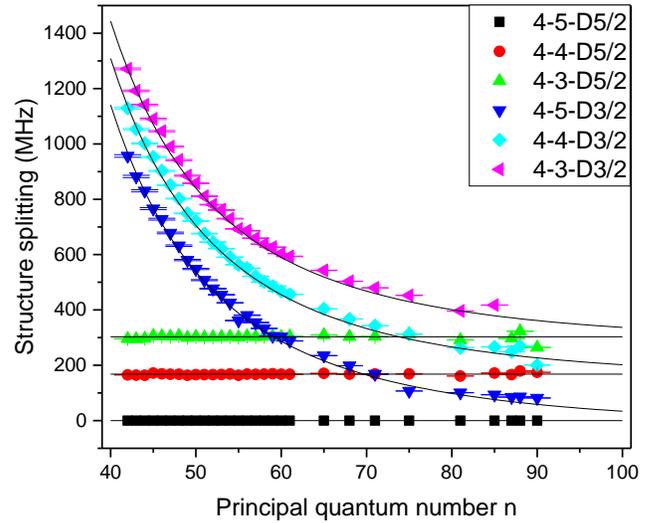

Fig. 4. Spectral splitting as a function of the principal quantum number $n$ from experimental values (color symbol) and theoretical results (black line). The errors are the percent of the bias errors between the theoretical and measured data.

Usually, it is difficult to observe the fine structure of the Rydberg states because the spectral intensity is very weak in the limit of the Boltzmann velocity distribution. The EIT signal intensity depends on the intensity and detuning of the laser, while the signal-to-noise ratio depends on the laser intensity noise and phase noise. To achieve a comparable spectral intensity, it is necessary to consider the effect arising from pumping the probe laser, including detuning and intensity. The origin of the transmittance signal can include the EIT, two-photon absorption (TPA), double-resonance optical pumping (DROP) and single-resonance optical pumping (SROP), for special experimental parameters. In the case of an open atomic configuration, the magnitudes of the transmittance spectra slowly decrease compared with that of a cycling transition owing to optical pumping. In the case of the closed ladder-type atomic system, EIT is more dominant than TPA under the conditions of a weak probe and a strong coupling laser power. The SROP and DROP are considered to be single- or two-photon resonance pumping and will spontaneously decay into other hyperfine states that depend on the detuning of the probing and coupling lasers, and are therefore weak under the condition of weak intensity and far detuning.

For our parameters, the power of the probe laser is on the order of hundreds of nanowatts, and the length of the glass cell is several millimeters to match the focused Rayleigh length, wherein the phase noise can be significantly suppressed. As the Doppler shift compensates the frequency detuning of the probe laser, atoms in a certain velocity group begin to resonate with the probe laser, whereupon the six EIT peaks with considerable signal-to-noise are observed. Owing to the Doppler effect, the probe laser resonant with $F=4 \rightarrow F'=5$ is a cycling transition that can suppress the SROP and DROP. For the larger absolute frequency detuning, we can achieve an optimized transmission signal and the linewidth exhibits a small increase, which signifies that the EIT signal is dominant.

We observe hyperfine-structure splitting of the intermediate states and fine-structure splitting of the Rydberg states. The measured results deviate from the theoretical results at higher *n* values, which may arise from the strong interaction between the highly-excited Rydberg atoms that induces energy state shifts. The interactions between the S and D states are repulsive or attractive depending on the special Rydberg states. Studies of the Rydberg atom interaction require extremely precise optical detection. Energy shifts arising from the van der Waals and dipole–dipole interactions, Forster resonance, etc., scale as tens of megahertz, and with the ~9 MHz-resolution bandwidth it is difficult to achieve accurate measurements at such frequency scales. Our experiments are limited by the noise arising from the laser phase noise, though implementation of the homodyne technique can effectively decrease the probe laser phase noise[19]. The crucial points are narrowing the linewidth of the optical detection by suppressing the intensity and phase noise.

In summary, we have demonstrated measurement of the fine structure of the Rydberg states by velocity-selective spectroscopy combined with RF bandwidth modulation. The experimental results show that this optical technique can be used to measure the Rydberg series of an energy level. The optical spectrum resolution is limited by the linewidth of the EIT signal, which can be improved by decreasing the laser phase noise in the condition of weak laser power. Velocity-selective spectroscopy combined with RF bandwidth modulation is an effective technique for studying interval spectrum splitting. It is not only an effective method for optical detection of the fine structure of Rydberg states, but it is also a good way to measure the strong interactions between Rydberg atoms at a weak laser power. Moreover, the non-destructive EIT detection demonstrated herein makes it possible to advance this technique to achieve high-precision measurements of electric fields and microwaves at the single-photon level.

**Funding.** This research work is financially supported by the National Natural Science Foundation of China (Grant Nos. 61205215, 61475091, and 61205215).

**Acknowledgment**. We thank the Professor Ken'ichi Nakagawa (Institute for Laser Science, University of Electro-Communications) for the helpful suggestions.

**References**

1. J. D. Pritchard, D. Maxwell, A. Gauguet, K. J. Weatherill, M. P. A. Jones, and C. S. Adams, Phys. Rev. Lett. **105,** 193603 (2010).
2. Y. O. Dudin and A. Kuzmich, Science **336,** 887-889 (2012).
3. T. Peyronel, O. Firstenberg, Q. Y. Liang, S. Hofferberth, A. V. Gorshkov, T. Pohl, M. D. Lukin, and V. Vuletić, Nature, **488,** 57-60 (2012).
4. D. Maxwell, D. J. Szwer, D. Paredes-Barato, H. Busche, J. D. Pritchard, A. Gauguet, K. J. Weatherill, M. P. A. Jones, and C. S. Adams., Phys. Rev. Lett. **110,** 103001 (2013).
5. M. Saffman, T. G. Walker, and K. Moelmer, Rev. Mod. Phys, **82,** 2313-2363 (2010).
6. D. Jaksch, J. I. Cirac, P. Zoller, S. L. Rolston, R. Cote, and M. D.Lukin, Phys. Rev. Lett. **85,** 2208 (2000).
7. M. D. Lukin, M. Fleischhauer, R. Cote, L. M. Duan, D. Jaksch, J. I. Cirac, and P. Zoller, Phys. Rev. Lett. **87,** 037901 (2001).
8. M. Fleischhauer, A. Imamoglu, and J. P. Marangos, Rev. Mod. Phys. **77,** 633-673 (2005).
9. A. K. Mohapatra, T. R. Jackson, and C. S. Adams, Phys. Rev. Lett. **98,** 113003 (2007).
10. M. Müller, I. Lesanovsky, H. Weimer, H. P. Buchler, and P. Zoller, Phys. Rev. Lett. **102,** 170502 (2009).
11. M. G. Bason, M. Tanasittikosol, A. Sargsyan, A. K. Mohapatra, D. Sarkisyan, R. M. Potvliege ,and C. S. Adams, New J. Phys. **12,** 065015 (2010).
12. D. Barredo, H. Kubler, R. Daschner, R. L¨ow, and T. Pfau, Phys. Rev. Lett. **110,** 123002 (2013).
13. H. Kübler, J. P. Shaffer, T. Baluktsian, R. Löw and T. Pfau, Nature Photonics **4,** 112-16 (2010).
14. S. A. Miller. D. A. Anderson, and G. Raithel, New J. Phys. **18,** 053017(2016).
15. Y. C. Jiao, X. X. Han, Z. W. Yang, J. K. Li, G. Raithel, J. M. Zhao, and S. T. Jia, Phys. Rev. A **94,** 023832 (2016).
16. W. Xu and B. DeMarco, Phys. Rev. A **93,** 011801(R) (2016).
17. S. X. Bao, H. Zhang, J. Zhou, L. J. Zhang, J. M. Zhao, L. T. Xiao, and S. T. Jia, Phys. Rev. A **94,** 043822 (2016).
18. A. Tauschinsky, R. Newell, H. B. van Linden van den Heuvell, and R. J. C. Preeuw, Phys. Rev. A **87,** 042522 (2013).
19. S. Kumar, H. Q. Fan, H. Kübler, A. J. Jahangiri, J. P. Shafferr, Sci Rep, **7,** 42981 (2017).